\newcommand{\ba}{\begin{array}}
\newcommand{\ea}{\end{array}}
\newcommand{\babc}{\begin{abc}}
\newcommand{\eabc}{\end{abc}}
\newcommand{\bc}{\begin{center}}
\newcommand{\ec}{\end{center}}
\newcommand{\be}{\begin{equation}}
\newcommand{\ee}{\end{equation}}
\newcommand{\bea}{\begin{eqnarray}}
\newcommand{\eea}{\end{eqnarray}}
\newcommand{\beas}{\begin{eqnarray*}}
\newcommand{\eeas}{\end{eqnarray*}}
\newcommand{\bh}{\begin{hangitem}}
\newcommand{\eh}{\end{hangitem}}
\newcommand{\bhi}{\begin{hangitem}}
\newcommand{\ehi}{\end{hangitem}}
\newcommand{\bi}{\begin{itemize}}
\newcommand{\ei}{\end{itemize}}
\newcommand{\bn}{\begin{enumerate}}
\newcommand{\en}{\end{enumerate}}
\newcommand{\bq}{\begin{quote}}
\newcommand{\eq}{\end{quote}}
\newcommand{\btb}{\begin{tabular}}
\newcommand{\etb}{\end{tabular}}
\def\litem[#1]{\item[#1\hfill]}         
\newtheorem{theorem}{Theorem}

\newtheorem{assumption}{Assumption}

\newtheorem{remark}{Remark}

\def\bfx{{\bold x}}
\def\bfy{{\bold y}}
\def\bfz{{\bold z}}
\documentclass[twocolumn]{autart}    

\usepackage{cite}
\usepackage{amsmath,amssymb,amsfonts}
\usepackage{algorithmic}
\usepackage{graphicx}
\usepackage{textcomp}
\usepackage{verbatim} 
\usepackage{epstopdf}
\usepackage{array}
\usepackage{arydshln}
\usepackage{amstext}
\usepackage{nccrules}
\usepackage{amsbsy}
\usepackage{mathptmx}
\usepackage{helvet}
\usepackage{courier}
\usepackage{type1cm}  
\usepackage{enumitem} 

\begin{document}

\begin{frontmatter}

\title{A Practical Application of Sliding Mode Control in the Motion Control of a High Precision Piezoelectric Motor} 

\thanks[footnoteinfo]{Corresponding author Khalid Abidi (e-mail: khalid.abidi@ncl.ac.uk).}

\author[Paestum]{Gangfeng~Yan}\ead{yangangfeng@cdu.edu.cn},    
\author[Rome]{and~Khalid~Abidi}\ead{khalid.abidi@ncl.ac.uk}             

\address[Paestum]{College of Information Science and Engineering, Chengdu University, ChengDu, SC 610106 CHINA}  
\address[Rome]{Newcastle University in Singapore, Electrical Power Engineering Program, SIT Building @ NYP, 172A Ang Mo Kio Avenue 8 \#05-01, Singapore 567739}             


\begin{abstract}                          
This paper proposes a practical implementation of sliding mode control (SMC) that utilizes partial modeling compensation. Sliding mode control is well known for its effectiveness as a model free control approach, however, its effectiveness is degraded if there is a constraint on the control gain or limitation on the switching frequency in digital implementation. This is especially the case with systems that involve static friction. This approach aims to enhance the effectiveness of SMC by partial model compensation. Rigorous stability proofs are presented to validate the approach. In addition, experiments are carried out on a piezoelectric motor driven linear stage and the control approach is compared with the Discrete-Time Integral Sliding Mode (DTISMC) approach proposed by \textit{Abidi et al.} as well as conventional PI control. The results show that the proposed control approach has a superior performance in comparison to the other approaches tested.
\end{abstract}

\end{frontmatter}

\section{Introduction}
Piezoelectric actuators are an attractive choice for high precision positioning applications that require sub-micrometer down to nanometer motion. The main characteristics of piezoelectric actuators are: quick response time, extremely high resolution in the nanometer range, high bandwidth, large force output, and a very short travel in the sub-millimeter range. Thus, piezoelectric actuators are ideal for very high-precision motion applications.  Application areas of piezoelectric actuators include: atomic force microscopes, adaptive optics, computer components, micromanipulation, micro-assembly, add-ons for high precision cutting machinery and as secondary actuators in macro/micro motion systems such as dual-stage hard-disk drives, \cite{chang1999,croft2000,song2009,yang2010}.

While piezoelectric actuators are extremely suitable for high precision control tasks, the nature of the control problem and the inherently nonlinear behavior of piezoelectric actuators means that the full capabilities of these actuators can only be realised with careful control. This, however, is challenging due to the complexity of the nonlinearity, which is a combination of hysteresis and creep phenomena. Nevertheless, many advanced control methods have been successfully applied to piezoelectric actuators (see \cite{xu2016}). The choice of method certainly depends on the application. For example, Iterative Learning Control is suitable for repetitive motions \cite{abidi2011}, while the robustness of Sliding Mode Control is effective in counteracting exogeneous disturbances as shown in \cite{Onal_Abidi,abidi_asif_07,abidi_xu2008,abidi2009,Ruixu2018,Yulong2017}.

Sliding Mode Control (SMC) is an effective control method for linear systems, nonlinear systems, time varying systems and uncertain systems, and is a powerful method for robust control \cite{Bar08}. In SMC, the controller is designed by choosing a suitable sliding mode surface based on the required closed-loop performance requirements. After the system states reach the sliding surface, under certain matching conditions, the system is said to be in a sliding mode regime and becomes completely robust or insensitive to exogenous disturbances. This characteristic of SMC makes it a superior choice of robust control. It is due to this unique characteristic that SMC is an attractive method for solving complex control problems and, hence, it is widely adopted in various types of industrial applications \cite{She14, Ma14, Sch15, Rau15, Muj15, Kwo14, Du_Chen_2018}.

Most SMC designs require full state information, which is a drawback due to the fact that only the output measurement is available in many practical applications as shown in \cite{abidi2009} and \cite{Xu08}. To solve this problem, some approaches required the design of state observers to construct the missing states, \cite{Gin14,Xu08,ref4}, while other approaches relied on the use of adaptive methods to compensate for the unknown state information, \cite{Fan16} and \cite{Bae16}. Although these methods have achieved good control performance, the controller designs are very complex.

It is well known that the actual system to be controlled and the mathematical model used for the controller design are always different in any practical control problem. The difference mainly comes from exogenous disturbances, uncertain system parameters and un-modeled dynamics.  However, if it is possible to couple the results of partial modeling to the simple design process of sliding mode controllers then it would be possible to formulate an approach with practical significance. Most of the approaches that involve piezoelectric motors, model the hysterisis while mostly ignoring all other nonlinear characteristics, \cite{Ruixu2018, Yulong2017}. In many of those approaches the critical aspects of the system (such as friction) are left unmodelled and are compensated using disturbance observers, \cite{Xu08, abidi2009}.  The drawback of using disturbance observers is the delay in reacting to the disturbance which results in less than optimal performance. In this paper, a simple design approach of SMC is proposed based on the partial modeling of the system. The aim of this work is to show an approach that greatly improves the performance of SMC based approaches in the presence of constraints on the control gain and switching frequency. Through experimental evaluations, the effectiveness of the proposed approach is confirmed.

The paper is organized as follows: The SMC approach with partial modeling compensation is introduced in Section II along with the stability analysis. In Section III, using the piezoelectric motor as a test bed, a detailed modeling process and the application of the control approach is presented in addition to comaparisons with other approaches. In Section V, conclusions are given.

Throughout this paper, for notational convenience, the mathematical expression ``$\dot{y}$'' represents the first derivative of $y$ with respect to time $t$ and ``$y^{(k)}$'' represents the $k^\mathrm{th}$ derivative of $y$ with respect to time $t$, respectively.

\section{SMC with Partial Modeling Compensation}
In this section, a problem statment that includes the general description of the system is given. The problem statement is then followed by a detailed derivation and stability analysis of the sliding mode control approach with partial modeling compensation (SMCPMC).
\subsection{Problem Statement}
Consider a system composed of $m$ scalar nonlinear ODEs written in the control normal form given by
\begin{equation}
\label{1}
\mathbf{y}^{(n)}(t) = \pmb{\xi}\left(\bar \bfx(t)\right)+\mathbf{d}(t)+\mathbf{u}(t)
\end{equation}
where $\bfy^\top(t) = \left [\begin{array}{ccccccc}\hspace{-0.15cm} y_1 &\hspace{-0.15cm}\mbox{\rotatebox[x=3pt,y=0pt]{90}{\dashrule{1.5 1 1.5 1 1.5 1 1.5 1 1.5}{.2}}}& \hspace{-0.15cm} y_2 &\hspace{-0.15cm} \mbox{\rotatebox[x=3pt,y=0pt]{90}{\dashrule{1.5 1 1.5 1 1.5 1 1.5 1 1.5}{.2}}}& \hspace{-0.15cm} \cdots & \hspace{-0.15cm} \mbox{\rotatebox[x=3pt,y=0pt]{90}{\dashrule{1.5 1 1.5 1 1.5 1 1.5 1 1.5}{.2}}}& \hspace{-0.15cm} y_m\end{array}\hspace{-0.15cm} \right] \in \mathcal{R}^m$ is the vector of outputs and $\mathbf{y}^{(n)}(t)=\left [\begin{array}{ccccccc}\hspace{-0.15cm} y_1^{(n_1)} &\hspace{-0.15cm}\mbox{\rotatebox[x=3pt,y=0pt]{90}{\dashrule{1.5 1 1.5 1 1.5 1 1.5 1 1.5}{.2}}}& \hspace{-0.15cm} y_2^{(n_2)} &\hspace{-0.15cm} \mbox{\rotatebox[x=3pt,y=0pt]{90}{\dashrule{1.5 1 1.5 1 1.5 1 1.5 1 1.5}{.2}}}& \hspace{-0.15cm} \cdots & \hspace{-0.15cm} \mbox{\rotatebox[x=3pt,y=0pt]{90}{\dashrule{1.5 1 1.5 1 1.5 1 1.5 1 1.5}{.2}}}& \hspace{-0.15cm} y_m^{(n_m)}\end{array}\hspace{-0.15cm} \right]^\top \in \mathcal{R}^m$ with $n_1, n_2, \cdots, n_m$ being the order of each scalar nonlinear ODE. The state vector for the $i^\mathrm{th}$ ODE is given as $\bfx_i^\top(t)=\left [\begin{array}{ccccccc}\hspace{-0.15cm} y_i &\hspace{-0.15cm}\mbox{\rotatebox[x=3pt,y=0pt]{90}{\dashrule{1.5 1 1.5 1 1.5 1 1.5 1 1.5}{.2}}}& \hspace{-0.15cm} \dot{y}_i &\hspace{-0.15cm} \mbox{\rotatebox[x=3pt,y=0pt]{90}{\dashrule{1.5 1 1.5 1 1.5 1 1.5 1 1.5}{.2}}}& \hspace{-0.15cm} \cdots & \hspace{-0.15cm} \mbox{\rotatebox[x=3pt,y=0pt]{90}{\dashrule{1.5 1 1.5 1 1.5 1 1.5 1 1.5}{.2}}}& \hspace{-0.15cm} y_i^{(n_i-1)}\end{array}\hspace{-0.15cm} \right]$ with the augmented state vector given as
${\bar \bfx}^\top(t)=\left [\begin{array}{ccccccc}\hspace{-0.15cm} \bfx_1^\top(t) &\hspace{-0.15cm}\mbox{\rotatebox[x=3pt,y=0pt]{90}{\dashrule{1.5 1 1.5 1 1.5 1 1.5 1 1.5}{.2}}}& \hspace{-0.15cm} \bfx_2^\top(t) &\hspace{-0.15cm} \mbox{\rotatebox[x=3pt,y=0pt]{90}{\dashrule{1.5 1 1.5 1 1.5 1 1.5 1 1.5}{.2}}}& \hspace{-0.15cm} \cdots & \hspace{-0.15cm} \mbox{\rotatebox[x=3pt,y=0pt]{90}{\dashrule{1.5 1 1.5 1 1.5 1 1.5 1 1.5}{.2}}}& \hspace{-0.15cm} \bfx_m^\top(t)\end{array}\hspace{-0.15cm} \right] \in \mathcal{R}^r$ where $r=\sum_{i=1}^m n_i$ and $i=1, 2, \cdots, m$. Finally, the vector of nonlinear functions is given as $\pmb{\xi}^\top(\cdot)=\left [\begin{array}{ccccccc}\hspace{-0.15cm} \xi_1(\cdot) &\hspace{-0.15cm}\mbox{\rotatebox[x=3pt,y=0pt]{90}{\dashrule{1.5 1 1.5 1 1.5 1 1.5 1 1.5}{.2}}}& \hspace{-0.15cm} \xi_2(\cdot) &\hspace{-0.15cm} \mbox{\rotatebox[x=3pt,y=0pt]{90}{\dashrule{1.5 1 1.5 1 1.5 1 1.5 1 1.5}{.2}}}& \hspace{-0.15cm} \cdots & \hspace{-0.15cm} \mbox{\rotatebox[x=3pt,y=0pt]{90}{\dashrule{1.5 1 1.5 1 1.5 1 1.5 1 1.5}{.2}}}& \hspace{-0.15cm} \xi_m(\cdot)\end{array}\hspace{-0.15cm} \right] \in \mathcal{R}^m$ while the vector of system disturbance is given as 
$\mathbf{d}^\top(t)=\left[\begin{array}{ccccccc}\hspace{-0.15cm} d_1(t) &\hspace{-0.15cm}\mbox{\rotatebox[x=3pt,y=0pt]{90}{\dashrule{1.5 1 1.5 1 1.5 1 1.5 1 1.5}{.2}}}& \hspace{-0.15cm} d_2(t) &\hspace{-0.15cm} \mbox{\rotatebox[x=3pt,y=0pt]{90}{\dashrule{1.5 1 1.5 1 1.5 1 1.5 1 1.5}{.2}}}& \hspace{-0.15cm} \cdots & \hspace{-0.15cm} \mbox{\rotatebox[x=3pt,y=0pt]{90}{\dashrule{1.5 1 1.5 1 1.5 1 1.5 1 1.5}{.2}}}& \hspace{-0.15cm} d_m(t)\end{array}\hspace{-0.15cm} \right] \in \mathcal{R}^m$ and the vector of control inputs is given as $\mathbf{u}^\top(t)=\left[\begin{array}{ccccccc}\hspace{-0.15cm} u_1(t) &\hspace{-0.15cm}\mbox{\rotatebox[x=3pt,y=0pt]{90}{\dashrule{1.5 1 1.5 1 1.5 1 1.5 1 1.5}{.2}}}& \hspace{-0.15cm} u_2(t) &\hspace{-0.15cm} \mbox{\rotatebox[x=3pt,y=0pt]{90}{\dashrule{1.5 1 1.5 1 1.5 1 1.5 1 1.5}{.2}}}& \hspace{-0.15cm} \cdots & \hspace{-0.15cm} \mbox{\rotatebox[x=3pt,y=0pt]{90}{\dashrule{1.5 1 1.5 1 1.5 1 1.5 1 1.5}{.2}}}& \hspace{-0.15cm} u_m(t)\end{array}\hspace{-0.15cm} \right] \in \mathcal{R}^m$, respectively.

\begin{assumption}
The nonlinear function $\xi_i(\cdot)$ is assumed uncertain and for any vectors $\bfz$ and $\mathbf{w}$, the nonlinear function is bounded as $|\xi_i(\bfz)-\xi_i(\mathbf{w})| \leq L_i\|\bfz - \mathbf{w}\|$ where $L_i$ is a positive constant and $\|\cdot\|$ is the Euclidean norm.
\end{assumption}

\begin{assumption}
The disturbance $\mathbf{d}(t)$ is not known \textit{a priori} and is bounded as $|d_i| \leq D_i$, where $D_i$ is a positive constant.
\end{assumption}

Consider now the system (\ref{1}), with the uncertainty assumptions on $\pmb{\xi}({\bar \bfx}(t))$ and $\mathbf{d}(t)$ a partial model of the system can be given as
\begin{equation}
\label{2}
\mathbf{y}^{(n)}(t) = \hat{\pmb{\xi}}({\bar \bfx}(t))+\mathbf{u}(t)
\end{equation}
where $\hat{\pmb{\xi}}({\bar \bfx}(t))$ is a partial model of the nonlinear function $\pmb{\xi}({\bar \bfx}(t))$ and the relationship between the partial model and the actual model is given by
\begin{equation}
\label{3}
\pmb{\xi}({\bar \bfx}(t))=\hat{\pmb{\xi}}({\bar \bfx}(t))+\tilde{\pmb{\xi}}({\bar \bfx}(t))
\end{equation}
with the term $\tilde{\pmb{\xi}}({\bar \bfx}(t))$ being the modeling error. The dimensions of $\hat{\pmb{\xi}}({\bar \bfx}(t))$ and $\tilde{\pmb{\xi}}({\bar \bfx}(t))$ are the same as that of $\pmb{\xi}({\bar \bfx}(t))$.

The control objective is to design a Sliding Mode Control law for the system (\ref{1}) compensated with the partial model (\ref{2}) such that stable and high-precision reference tracking is achieved. The controller design and stability analysis is presented in the next section.

\subsection{Derivation of the Control Approach}
To proceed with the design of the SMCPMC controller, the tracking error is defined as
\begin{equation}
\label{4}
e_i(t) = y_{\mathrm{d},i}(t) - y_i(t)
\end{equation}
where $y_{\mathrm{d},i}$ denotes $i^\mathrm{th}$ desired reference trajectory. For the sake of convenience, $t$ will be omitted for the remainder of this section. Consider now, the sliding surface given as
\begin{equation}
\label{5}
s_i(\mathbf{e}_i)=\sum_{k=0}^{n_i-1}\lambda_{i,k}e_i^{(k)}
\end{equation}
where $\mathbf{e}_i^\top=\left[\begin{array}{ccccccc}\hspace{-0.15cm} e_i &\hspace{-0.15cm}\mbox{\rotatebox[x=3pt,y=0pt]{90}{\dashrule{1.5 1 1.5 1 1.5 1 1.5 1 1.5}{.2}}}& \hspace{-0.15cm} \dot{e}_i &\hspace{-0.15cm} \mbox{\rotatebox[x=3pt,y=0pt]{90}{\dashrule{1.5 1 1.5 1 1.5 1 1.5 1 1.5}{.2}}}& \hspace{-0.15cm} \cdots & \hspace{-0.15cm} \mbox{\rotatebox[x=3pt,y=0pt]{90}{\dashrule{1.5 1 1.5 1 1.5 1 1.5 1 1.5}{.2}}}& \hspace{-0.15cm} e_i^{(n_i-1)}\end{array}\hspace{-0.15cm} \right]$ and $\lambda_{i,k}$ are positive constants to be designed based on the performance characteristics required.

\begin{theorem}
For the system described by (\ref{1}) and the sliding surface (\ref{5}), the tracking error converges to zero asymptotically if the control law is selected as
\begin{equation}
\label{6}
\mathbf{u}=\hat{\mathbf{u}}+\pmb{\varphi} +\mathbf{\eta}\circ\mathbf{s}+\pmb{\beta}\circ \mathrm{sgn}(\mathbf{s})
\end{equation}
where $\hat{\mathbf{u}}^\top=\left[\begin{array}{ccccccc}\hspace{-0.15cm} \hat{u}_1 &\hspace{-0.15cm}\mbox{\rotatebox[x=3pt,y=0pt]{90}{\dashrule{1.5 1 1.5 1 1.5 1 1.5 1 1.5}{.2}}}& \hspace{-0.15cm} \hat{u}_2 &\hspace{-0.15cm} \mbox{\rotatebox[x=3pt,y=0pt]{90}{\dashrule{1.5 1 1.5 1 1.5 1 1.5 1 1.5}{.2}}}& \hspace{-0.15cm} \cdots & \hspace{-0.15cm} \mbox{\rotatebox[x=3pt,y=0pt]{90}{\dashrule{1.5 1 1.5 1 1.5 1 1.5 1 1.5}{.2}}}& \hspace{-0.15cm} \hat{u}_m\end{array}\hspace{-0.15cm} \right]$ is the component of the control law corresponding to the inverse of the partial model and
\begin{equation}
\label{6-1}
\pmb{\varphi}=\left[\begin{array}{cccc}\frac{1}{\lambda_{1,n_1}}\sum_{k=1}^{n_1-1}\lambda_{1,k}e_1^{(k)} \\
\frac{1}{\lambda_{2,n_2}}\sum_{k=1}^{n_2-1}\lambda_{2,k}e_2^{(k)} \\ \vdots \\
\frac{1}{\lambda_{m,n_m}}\sum_{k=1}^{n_m-1}\lambda_{m,k}e_m^{(k)}\end{array}\right].
\end{equation}
Furthermore, the vectors $\pmb{\eta}^\top=\left[\begin{array}{ccccccc}\hspace{-0.15cm} \eta_1 &\hspace{-0.15cm}\mbox{\rotatebox[x=3pt,y=0pt]{90}{\dashrule{1.5 1 1.5 1 1.5 1 1.5 1 1.5}{.2}}}& \hspace{-0.15cm} \eta_2 &\hspace{-0.15cm} \mbox{\rotatebox[x=3pt,y=0pt]{90}{\dashrule{1.5 1 1.5 1 1.5 1 1.5 1 1.5}{.2}}}& \hspace{-0.15cm} \cdots & \hspace{-0.15cm} \mbox{\rotatebox[x=3pt,y=0pt]{90}{\dashrule{1.5 1 1.5 1 1.5 1 1.5 1 1.5}{.2}}}& \hspace{-0.15cm} \eta_m\end{array}\hspace{-0.15cm} \right] \in \mathcal{R}^m$, $\pmb{\beta}^\top=\left[\begin{array}{ccccccc}\hspace{-0.15cm} \beta_1 &\hspace{-0.15cm}\mbox{\rotatebox[x=3pt,y=0pt]{90}{\dashrule{1.5 1 1.5 1 1.5 1 1.5 1 1.5}{.2}}}& \hspace{-0.15cm} \beta_2 &\hspace{-0.15cm} \mbox{\rotatebox[x=3pt,y=0pt]{90}{\dashrule{1.5 1 1.5 1 1.5 1 1.5 1 1.5}{.2}}}& \hspace{-0.15cm} \cdots & \hspace{-0.15cm} \mbox{\rotatebox[x=3pt,y=0pt]{90}{\dashrule{1.5 1 1.5 1 1.5 1 1.5 1 1.5}{.2}}}& \hspace{-0.15cm} \beta_m\end{array}\hspace{-0.15cm} \right]  \in \mathcal{R}^m$ are positive constants and $\mathbf{s}^\top=\left[\begin{array}{ccccccc}\hspace{-0.15cm} s_1(\mathbf{e}_1) &\hspace{-0.15cm}\mbox{\rotatebox[x=3pt,y=0pt]{90}{\dashrule{1.5 1 1.5 1 1.5 1 1.5 1 1.5}{.2}}}& \hspace{-0.15cm} s_2(\mathbf{e}_2) &\hspace{-0.15cm} \mbox{\rotatebox[x=3pt,y=0pt]{90}{\dashrule{1.5 1 1.5 1 1.5 1 1.5 1 1.5}{.2}}}& \hspace{-0.15cm} \cdots & \hspace{-0.15cm} \mbox{\rotatebox[x=3pt,y=0pt]{90}{\dashrule{1.5 1 1.5 1 1.5 1 1.5 1 1.5}{.2}}}& \hspace{-0.15cm} s_m(\mathbf{e}_m)\end{array}\hspace{-0.15cm} \right] \in \mathcal{R}^m$ while `$\circ$' denotes the Schur product and $\mathrm{sgn}(\cdot)$ denotes the signum function.  The values of $\mathbf{\eta}$ and $\mathbf{\beta}$ can be tuned by trial and error observation.
\end{theorem}

\textit{Proof}:
To demonstrate the stability of the designed control approach, consider a positive Lyapunov function defined as
\begin{equation}
\label{7}
V=\frac{1}{2}\mathbf{s^\top s}.
\end{equation}
Differentiating (\ref{7}) with respect to time, the time derivative of the Lyapunov function is obtained as
\begin{equation}
\label{8}
\dot{V}=\mathbf{s}^\top \mathbf{\dot{s}}
\end{equation}
where $\mathbf{\dot{s}}^\top=\left[\begin{array}{ccccccc}\hspace{-0.15cm} \dot{s}_1(\mathbf{e}_1) &\hspace{-0.15cm}\mbox{\rotatebox[x=3pt,y=0pt]{90}{\dashrule{1.5 1 1.5 1 1.5 1 1.5 1 1.5}{.2}}}& \hspace{-0.15cm} \dot{s}_2(\mathbf{e}_2) &\hspace{-0.15cm} \mbox{\rotatebox[x=3pt,y=0pt]{90}{\dashrule{1.5 1 1.5 1 1.5 1 1.5 1 1.5}{.2}}}& \hspace{-0.15cm} \cdots & \hspace{-0.15cm} \mbox{\rotatebox[x=3pt,y=0pt]{90}{\dashrule{1.5 1 1.5 1 1.5 1 1.5 1 1.5}{.2}}}& \hspace{-0.15cm} \dot{s}_m(\mathbf{e}_m)\end{array}\hspace{-0.15cm} \right]$. From the definitions of $\mathbf{s}$ and $\mathbf{\dot{s}}$ the time derivative of the Lyapunov function can be exapanded as
\begin{equation}
\label{10}
\dot{V}=\sum_i^m s_i(\mathbf{e}_i)\dot{s}_i(\mathbf{e}_i).
\end{equation}
In order to guarantee that $\dot{V}$ is negative definite, all the terms in $\sum_i^m s_i(\mathbf{e}_i)\dot{s}_i(\mathbf{e}_i)$ must be negative definite. However, due to the fact that the states $x_i$ are $m$ independent coordinates, it is sufficient to show that the $i^\mathrm{th}$ term $s_i(\mathbf{e}_i)\dot{s}_i(\mathbf{e}_i)$ is negative definite. In the following discussion, it will be shown that $s_i(\mathbf{e}_i)\dot{s}_i(\mathbf{e}_i)$ is negative definite.

Consider the time derivative of the $i^\mathrm{th}$ sliding surface (\ref{5}) given as
\begin{equation}
\label{9}
\dot{s}_i(\mathbf{e}_i)=\sum_{k=1}^{n_i}\lambda_{i,k}e_i^{(k)}
\end{equation}
then it is obtained that
\bea
s_i(\mathbf{e}_i)\dot{s}_i(\mathbf{e}_i)&=&s_i(\mathbf{e}_i)\sum_{k=1}^{n_i}\lambda_{i,k}e_i^{(k)}\nonumber\\
&=&s_i(\mathbf{e}_i)\left(\lambda_{i,n_i}e_i^{(n_i)}+\sum_{k=1}^{n_i-1}\lambda_{i,k}e_i^{(k)}\right).
\label{11}
\eea
Consider now the definition of the tracking error, substitution of the system (\ref{1}) into the $n_i^\mathrm{th}$ time derivative of the tracking error results in 
\bea
e_i^{(n_i)}&=&y_{\mathrm{d},i}^{(n_i)}-y_{i}^{(n_i)}\nonumber\\
&=&y_{\mathrm{d},i}^{(n_i)}-\xi_i({\bar \bfx}_i) - d_i - u_i\nonumber\\
&=&\hat{u}_i - \tilde{\xi}_i(\bar \bfx_{\mathrm{d},i}) - d_i - (\xi_i(\bar \bfx_{i})-\xi_i(\bar \bfx_{\mathrm{d},i}))-u_i
\label{12}
\eea
where ${\bar \bfx_{\mathrm{d},i}}^\top(t)=\left [\begin{array}{ccccccc}\hspace{-0.15cm} \bfx_{\mathrm{d},1}^\top(t) &\hspace{-0.15cm}\mbox{\rotatebox[x=3pt,y=0pt]{90}{\dashrule{1.5 1 1.5 1 1.5 1 1.5 1 1.5}{.2}}}& \hspace{-0.15cm} \bfx_{\mathrm{d},2}^\top(t) &\hspace{-0.15cm} \mbox{\rotatebox[x=3pt,y=0pt]{90}{\dashrule{1.5 1 1.5 1 1.5 1 1.5 1 1.5}{.2}}}& \hspace{-0.15cm} \cdots & \hspace{-0.15cm} \mbox{\rotatebox[x=3pt,y=0pt]{90}{\dashrule{1.5 1 1.5 1 1.5 1 1.5 1 1.5}{.2}}}& \hspace{-0.15cm} \bfx_{\mathrm{d},m}^\top(t)\end{array}\hspace{-0.15cm} \right]$ and $\bfx_{\mathrm{d},i}^\top(t)=\left [\begin{array}{ccccccc}\hspace{-0.15cm} y_{\mathrm{d},i} &\hspace{-0.15cm}\mbox{\rotatebox[x=3pt,y=0pt]{90}{\dashrule{1.5 1 1.5 1 1.5 1 1.5 1 1.5}{.2}}}& \hspace{-0.15cm} \dot{y}_{\mathrm{d},i} &\hspace{-0.15cm} \mbox{\rotatebox[x=3pt,y=0pt]{90}{\dashrule{1.5 1 1.5 1 1.5 1 1.5 1 1.5}{.2}}}& \hspace{-0.15cm} \cdots & \hspace{-0.15cm} \mbox{\rotatebox[x=3pt,y=0pt]{90}{\dashrule{1.5 1 1.5 1 1.5 1 1.5 1 1.5}{.2}}}& \hspace{-0.15cm} y_{\mathrm{d},i}^{(n_i-1)}\end{array}\hspace{-0.15cm} \right]$. If the terms $\tilde{\xi}_i(\bar \bfx_{\mathrm{d},i})$ , $d_i$ and $\xi_i(\bar \bfx_{i})-\xi_i(\bar \bfx_{\mathrm{d},i})$ can be compensated by proper estimates then it is guaranteed that $s_i(\mathbf{e}_i)\dot{s}_i(\mathbf{e}_i)$ is negative definite. Using \textbf{Assumption 1}, $\xi_i(\cdot)$ is bounded in a limited interval for the actual physical system, then $|\xi_i(\bar \bfx_{i})-\xi_i(\bar \bfx_{\mathrm{d},i})| \leq L_i\|\mathbf{e}_i\|$, where $L_i$ is a positive constant and $|\tilde{\xi}_i(\bar \bfx_{\mathrm{d},i})| \leq \rho_\mathrm{c}$ where $\rho_\mathrm{c}$ is a positive constant. The disturbance $d_i$ is assumed to be bounded as $|d_i| \leq D_i$, where $D_i$ is a positive constant. The upperbounds $\rho_\mathrm{c}$ and $D_i$ can be obtained by open-loop experiments before the implementation of the controller. If the coefficients $\eta_i$ and $\beta_i$ are selected properly, it is obtained that
\begin{equation}
\label{13}
\eta_i|s(\mathbf{e}_i)| + \beta_i\geq D_i+L_i\|\mathbf{e}_i\|+\rho_\mathrm{c}
\end{equation}
where $\eta_i$ and $\beta_i$ are selected to ensure that (\ref{13}) is satisfied. From the control law (\ref{6}), after performing some simplifications, it is obtained that
\begin{equation}
\label{14}
s_i(\mathbf{e}_i)\dot{s}_i(\mathbf{e}_i)\leq{0}.
\end{equation}
Therefore, the time derivative of Lyapunov function (\ref{8}) is negative definite. Further, according to the invariant set theorem, the system (\ref{1}) with control approach (\ref{6}) is asymptotically stable.

\begin{remark} 
From the condition (\ref{13}), it can be seen that swithing gain $\beta_i$ can be selected properly to ensure stability. However, in applications with limited switching frequency a large switching gain can lead to chattering. Thus, it is necessary to have a high accuracy system model in order to require a lower switching gain and a good system performance.
\end{remark}

\begin{remark} 
In the actual design of the SMCPMC control approach, due to the existence of various types of perturbations, the signal $e_i^{(k)}$ for $i=1,2,\cdots,m$ and  $k=0,1,\cdots,n_i-1$ may need to be filtered.
\end{remark}

\begin{figure}
\begin{center}
\includegraphics[width=7.5cm]{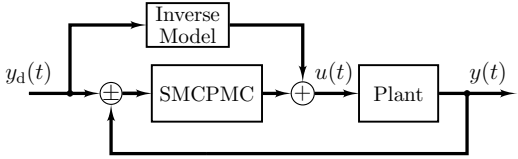}   
\caption{Block diagram of the proposed control approach.}  
\label{fig_sim_1}                                 
\end{center}                                 
\end{figure}

A block diagram that describes the implementation of the proposed control approach is shown in Fig. \ref{fig_sim_1}. This implementation will be tested experimentatlly in order to verify the effectiveness of the proposed control approach.


\section{Experimental Implementation}

\begin{figure}[!t]
\centering
\includegraphics[width=9.0cm]{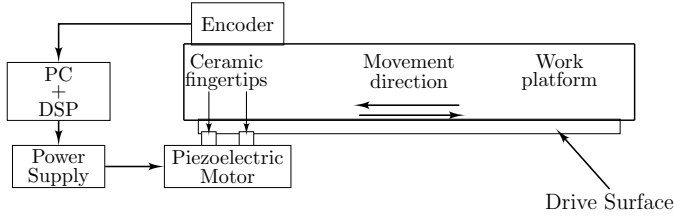}
\caption{Schematic illustrating the experimental setup.}
\label{fig:workingPrinciple}
\end{figure}

In this section the SMCPMC based control law is implemented on a piezoelectric motor driven linear stage manufactured by PBA Systems. The linear stage has a maximum range of travel of 115mm, and a maximum velocity of 230$\mathrm{mms}^{-1}$. The stage is actuated by a Nanomotion HR-8 piezoelectric motor. Its working principle can briefly described using Fig. \ref{fig:workingPrinciple}.
The actuating elements are a set of piezoelectric ceramic fingers. The fingertips, protruding from one end of the motor, are mounted in compression against the drive belt of the work platform. When driven by electrical signals from the motor driver, ultrasonic standing waves are produced and the high frequency longitudinal extension and lateral bending of the finger generates an elliptical motion at the fingertips. The force exerted on the drive belt by the fingertips moving in such a manner produces linear motion along the direction as shown in Fig. \ref{fig:workingPrinciple}. The control voltage applied to the motor driver determines the velocity of motion. In the absence of a drive voltage input, the pressure of the ceramic fingertips on the drive belt maintains a seizing force on the work platform.

\begin{figure}[!b]
\centering
\includegraphics[width=9.0cm]{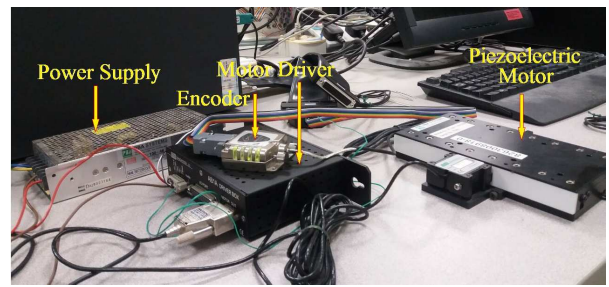}
\caption{Piezoelectric motor stage control system.}
\label{fig_sim_4}
\end{figure}

Discplacement is measured by a Mercury 3000 optical encoder made by Celera Motion while the velocity and acceleration signals are obtained by numerical differentiation of the position.
All control and measurement algorithms are implemented with MATLAB/SIMULINK on a host computer, and executed by a dSPACE DS1104 card installed inside. Signal acquisition and generation are respectively via the DS1104's 12-bit Analog-to-Digital Converter (ADC) channels (800ns conversion time) and 16-bit Digital-to-Analog converter (DAC) channels (10$\mu$s settling time), both having $\pm$10V dynamic range. These channels interface with the piezoelectric motor driver and the encoder. Through a user interface on the dSPACE ControlDesk software, experiments are performed with parameter adjustments and measurements made in real time. The system is shown in Fig. \ref{fig_sim_4}.

\subsection{Type of Friction Force}

A piezoelectric motor driven linear stage consists of a platform that slides on rigid rails and, as such, friction plays a major role in the disturbance that effects the performance of the system. Therefore, various friction force models will be discussed and that will be followed by an attempt to identify the friction model using an open-loop test to determine the system model parameters.

\subsubsection{Static Friction}
The static friction resists all motion as long as the driving force is smaller in magnitude than the maximum static friction force $f_\mathrm{s}$ at zero velocity. Static friction is discontinuous when the velocity crosses zero.

Static friction is described by
\begin{equation}
\label{21}
F_\mathrm{s}=\left\{ \begin{array}{lll}
0, &  |f_\mathrm{a}| \geq f_\mathrm{s} \; \mathrm{and} \; v\neq0 \\
f_\mathrm{a}, &  |f_\mathrm{a}|<f_\mathrm{s} \; \mathrm{and} \; v = 0 \\
f_\mathrm{s} \mathrm{sgn}(f_\mathrm{a}), & |f_\mathrm{a}| \geq f_\mathrm{s} \; \mathrm{and} \;  v = 0
\end{array} \right.
\end{equation}
where $F_\mathrm{s}$ is the static friction force, $f_\mathrm{a}$ is the applied force and $v = \dot{y}$ is the velocity.

\subsubsection{Coulomb Friction}
Coulomb friction is a type of mechanical damping in which energy is consumed via sliding friction. The friction generated by the relative motion of the two surfaces that press against each other always resists relative motion and is proportional to the normal force of contact. Coulomb friction is described by
\begin{equation}
\label{22}
F_\mathrm{c} = f_\mathrm{c} \mathrm{sgn}(v)
\end{equation}
where $F_\mathrm{c}$ is the Coulomb friction force and $f_\mathrm{c}$ is the normal force applied to the surface.

\subsubsection{Viscous Friction}
Viscous friction, is a resistance force that acts on an object in motion. Under well-lubricated conditions the viscous friction force is approximately proportional to velocity. It satisfies the linear relationship given as
\begin{equation}
\label{23}
F_\mathrm{v} = f_\mathrm{v} v
\end{equation}
where $F_\mathrm{v}$ is the viscous friction force and $f_\mathrm{v}$ is the coefficient of viscous friction.

\subsubsection{Drag Friction}
Drag friction is the friction force between a solid object and a liquid or a gas. It is proportional to the square of velocity and is described by
\begin{equation}
\label{24}
F_\mathrm{d}=f_\mathrm{d}v|v|
\end{equation}
where $F_\mathrm{d}$ is the drag friction force and $f_\mathrm{d}$ is the drag coefficient.

Classical friction models have different combinations of static, coulomb, viscous and drag friction as their basic components.

\subsection{System Modeling}
\begin{figure}[!t]
\centering
\includegraphics[width=8.4cm]{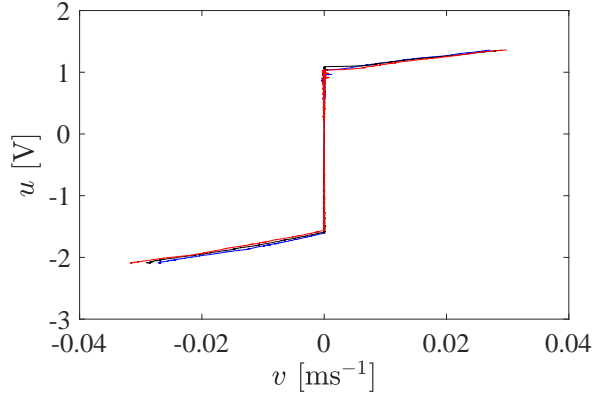}
\caption{Experimental results of control input $u$ w.r.t. velocity $v$.}
\label{fig_sim_5}
\end{figure}

A number of experiments are carried out, and the results of three experiments are shown in Fig. \ref{fig_sim_5}. In the experiment, a slow triangular input is used on the piezoelectric motor stage to generate a low velocity motion with low acceleration. This way, the input force is used solely to overcome the friction force of the piezoelectric motor stage. Thus, the force-velocity relationship in Fig. \ref{fig_sim_5} can be obtained. It can be seen that the static friction force, the Coulomb friction force and the viscous friction force models need to be considered for modeling the piezoelectric motor stage. The speed of piezoelectric motor stage is very low, and the coefficient of drag friction force is very low too, therefore, drag friction force can be neglected in this case.

\begin{figure}[!b]
\centering
\includegraphics[width=8.4cm]{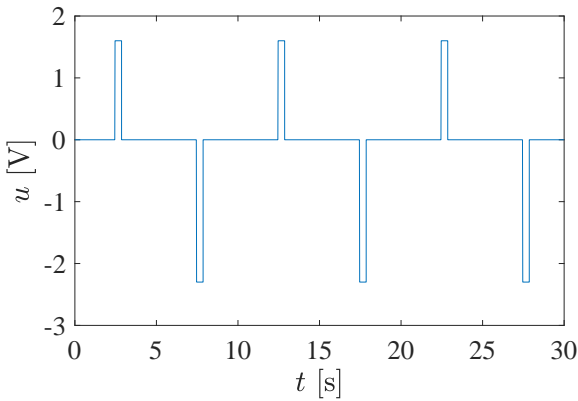}
\caption{Input pulse of 0.4s and -2.3V, 1.6V peak to peak.}
\label{fig_sim_6}
\end{figure}

\begin{figure}[!t]
\centering
\includegraphics[width=8.4cm]{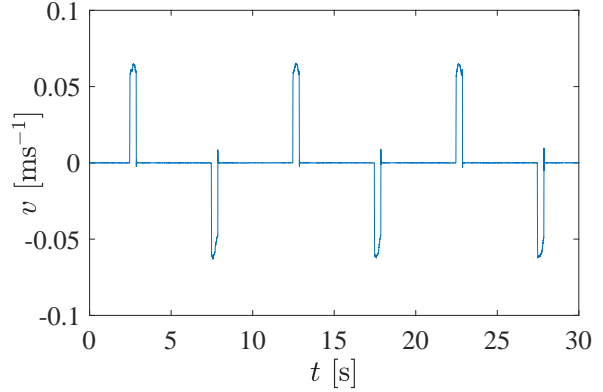}
\caption{Velocity response to of 0.4s pulse with -2.3V, 1.6V peak to peak.}
\label{fig_sim_7}
\end{figure}

In consideration of the static friction force, the Coulomb friction force and the viscous friction force, the dynamics of the piezoelectric motor can be represented by the following second-order differential equation according to Newton's second law
\begin{equation}
\label{25}
\dot{v}(t) = - \alpha_1 v(t) - \alpha_2\mathrm{sgn}(v(t)) - \alpha_\mathrm{s}\delta(v(t))\mathrm{sgn}(u(t)) + \alpha_3u(t)
\end{equation}
where $y(t)$ is the linear displacement, $v(t) = \dot{y}(t)$, $u(t)$ is the voltage input, $\alpha_1$ is the coefficients of viscous friction, $\alpha_2$ is the coefficients of Coulomb's friction, $\alpha_\mathrm{s}$ is the coefficients of static friction, $\alpha_3$ is force coefficients of voltage to force conversion and $\delta(v(t))$ is given as
\begin{equation}
\label{25-1}
\delta(v(t))=\left\{ \begin{array}{ll}
1, & \quad  v(t)= 0 \\
0, & \quad   v(t)\neq 0
\end{array} \right..
\end{equation}
Through experiments of the velocity response to an input $u(t)$ in the form of a triangle function as shown in Fig. \ref{fig_sim_5}, it is observed that the values $\alpha_1$ and $\alpha_2$ are not the same when the direction of the velocity is changed. Therefore, the model (\ref{25}) can be modified where $\alpha_1$ and $\alpha_2$ take different values for the different directions of $v(t)$ as
\begin{equation}
\label{25-2}
\alpha_1=\left\{ \begin{array}{ll}
\alpha_\mathrm{1p}, & \quad  v(t) > 0 \\
\alpha_\mathrm{1n}, & \quad  v(t) < 0
\end{array} \right. 
\end{equation}
and
\begin{equation}
\label{25-3}
\alpha_2=\left\{ \begin{array}{ll}
\alpha_\mathrm{2p}, & v(t) > 0 \\
\alpha_\mathrm{2n}, & v(t) <0
\end{array} \right.
\end{equation}
where $\alpha_\mathrm{1p}$ and $\alpha_\mathrm{2p}$ are the values of $\alpha_1$ and $\alpha_2$ that correspond to positive velocity direction whereas $\alpha_\mathrm{1n}$ and $\alpha_\mathrm{2n}$ are the values of $\alpha_1$ and $\alpha_2$ that correspond to negative velocity direction. The coefficient $\alpha_3 = 6 \mathrm{NV}^{-1}$ is provided in the piezoelectric motor product documentation. In order to obtain the values of the remaining parameters in model (\ref{25}), the system will be subjected to pulse inputs in order minimize the influence of static friction on the system. By using a pulse input of $0.4$s duration and with  $-2.3\mathrm{V}$, $1.6\mathrm{V}$ of amplitude, as shown in Fig. \ref{fig_sim_6}, the velocity response to the pulse inputs is shown in Fig. \ref{fig_sim_7}.
From the results, it is obtained that
\begin{equation}
\label{27}
\left\{ \begin{aligned}
&    0=0.05562\alpha_\mathrm{1n}+\alpha_\mathrm{2n}-2.3\alpha_3
\\&  0=-0.06222\alpha_\mathrm{1p}-\alpha_\mathrm{2p}+1.6\alpha_3
\end{aligned} \right.
\end{equation}
Following a similar method, using a pulse of $0.4$s duration and amplitudes of $-1.8$V, $1.3$V, $-2$V, $1.5$V, $-2.1$V, $1.7$V, $-2.5$V and $2$V respectively, it is obtained that
\begin{equation}
\label{28}
\left\{ \begin{aligned}
& 0 = 0.03393\alpha_\mathrm{1n} + \alpha_\mathrm{2n} - 1.8\alpha_3
\\&0 = 0.04622\alpha_\mathrm{1n} + \alpha_\mathrm{2n} - 2\alpha_3
\\&0 = 0.04991\alpha_\mathrm{1n} + \alpha_\mathrm{2n} - 2.1\alpha_3
\\&0 = 0.07120\alpha_\mathrm{1n} + \alpha_\mathrm{2n} - 2.5\alpha_3
\\&0 =  - 0.04465\alpha_\mathrm{1p} - \alpha_\mathrm{2p} + 1.3\alpha_3
\\&0 =  - 0.05742\alpha_\mathrm{1p} - \alpha_\mathrm{2p} + 1.5\alpha_3
\\&0 =  - 0.06863\alpha_\mathrm{1p} - \alpha_\mathrm{2p} + 1.7\alpha_3
\\&0 =  - 0.08519\alpha_\mathrm{1p} - \alpha_\mathrm{2p} + 2.0\alpha_3
\end{aligned} \right.
\end{equation}
%
%
Let $A^\top=[\alpha_\mathrm{1p},\alpha_\mathrm{1n},\alpha_\mathrm{2p},\alpha_\mathrm{2n}]$, $Y^\top=6\times[1.8,2,2.1,2.3, 2.5,$ $-1.3, -1.5, -1.6, -1.7, -2.0]$ and $X$ be the coefficients of $A$. Using the least-squares method given by
\begin{equation}
\label{29}
A=(X^\top X)^{-1}X^\top Y
\end{equation}
the coefficients $\alpha_\mathrm{1p}$, $\alpha_\mathrm{1n}$, $\alpha_\mathrm{2p}$ and $\alpha_\mathrm{2n}$ can be obtained, by solving the equation (\ref{29}), as
\\
$\\ 
\alpha_1=\left\{ \begin{aligned}
&         104.0154, \quad  v(t)> 0
\\& 117.1441, \quad   v(t)<0
\end{aligned} \right. \quad \quad
\alpha_2=\left\{ \begin{aligned}
&         3.1023, \quad  v(t)> 0
\\& 6.8216, \quad   v(t)<0
\end{aligned} \right.$
\vspace{0.1cm}

\begin{figure}[!b]
\centering
\includegraphics[width=8.4cm]{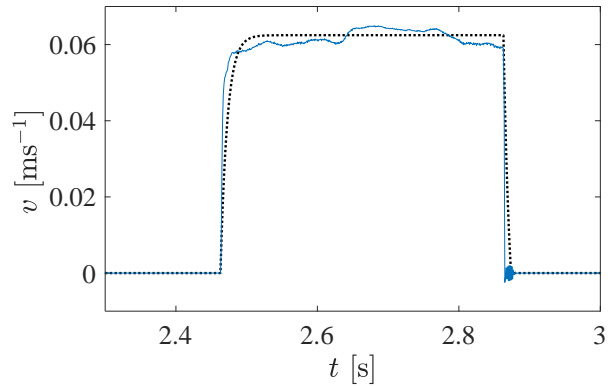}
\caption{Measured (solid) and modelled (dotted) velocity response to 0.4s pulse input with -2.3V, 1.6V peak to peak.}
\label{fig_sim_8_1}
\end{figure}

\begin{figure}[!t]
\centering
\includegraphics[width=8.4cm]{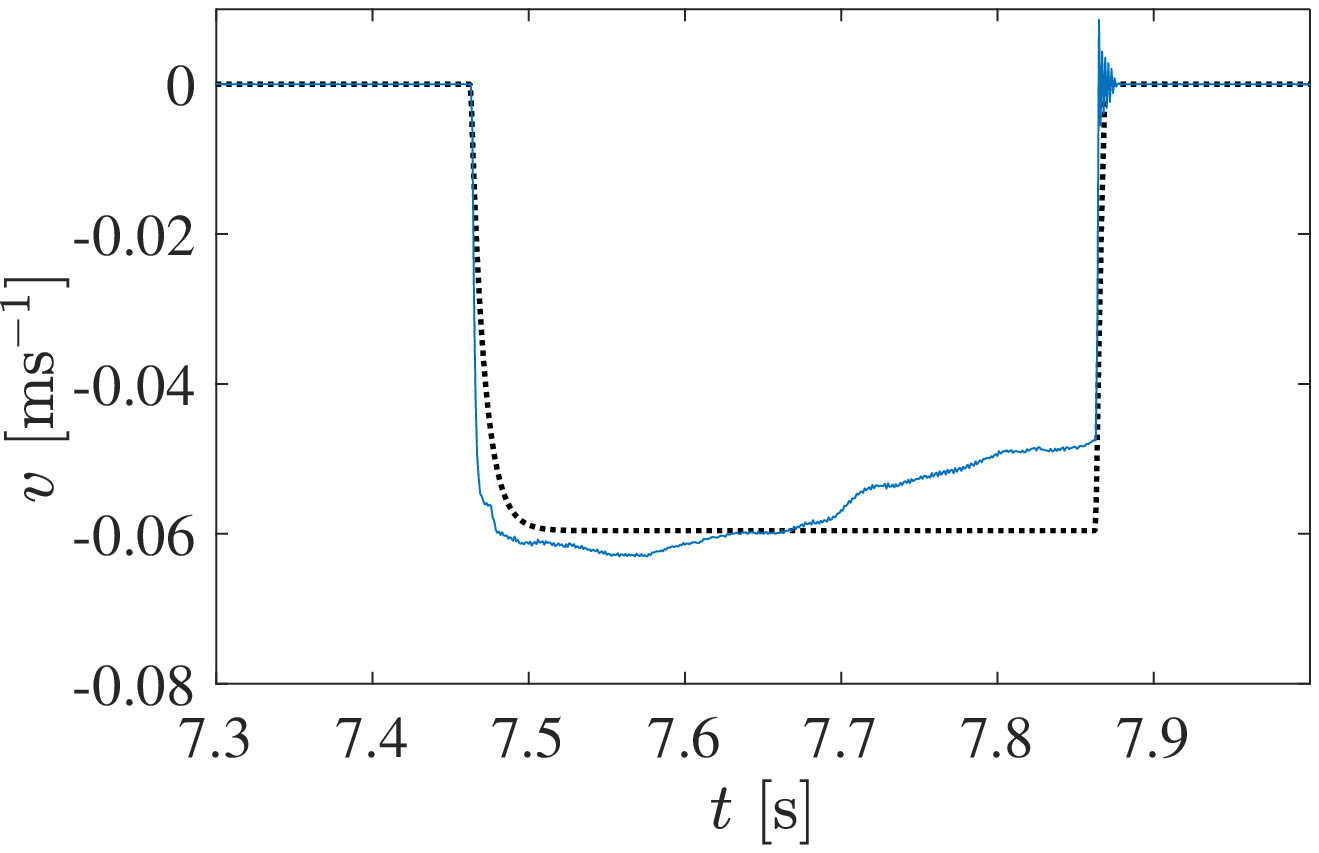}
\caption{Measured (solid) and modelled (dotted) velocity response to 0.4s pulse input with -2.3V, 1.6V peak to peak.}
\label{fig_sim_8_2}
\end{figure}

\hspace{-0.35cm}Using model (\ref{25}), while ignoring static friction, contrasting curves for velocities are obtained as a response to the pulses of amplitude $1.6V$ and $-2.3V$, as shown in Fig. \ref{fig_sim_8_1} and Fig. \ref{fig_sim_8_2}. In Fig. \ref{fig_sim_8_1} and Fig. \ref{fig_sim_8_2}, the dotted line shows the response of the model while the solid line shows the response of the actual system.
From Fig. \ref{fig_sim_8_1} and Fig. \ref{fig_sim_8_2}, it was observed that the viscous friction force in model (\ref{25}) has a certain delay, and considering the results of the triangular function input test, the model for the piezoelectric motor stage can be obtained using the performance analysis as follows
\begin{equation}  
\label{30}
\dot{v}(t)=\left\{ \begin{array}{lll}
-\alpha_1 v(t-\tau) - 3.1023 + 6u,   & v(t)> v_\mathrm{cr} \\
-\alpha_1 v(t-\tau) - \alpha_\mathrm{s} \mathrm{sgn}(v(t)) + 6u,  & |v(t)|\leq v_\mathrm{cr} \\
-\alpha_1 v(t-\tau) + \alpha_\mathrm{2n} + 6u,  & v(t)< -v_\mathrm{cr}
\end{array} \right.
\end{equation}
and
\\
\\
$\alpha_1=\left\{ \begin{aligned}
&    104.0154, \quad  v(t-\tau)> 0
\\& 117.1441, \quad   v(t-\tau)<0
\end{aligned} \right.
\\
\vspace{0.35cm}
\\
\alpha_\mathrm{2n}=\left\{ \begin{aligned}
&         5.8216 + \left(1-e^{-30v(t)}\right), \quad  \dot{u}(t)\leq 0
\\& 6.8216, \quad \quad\quad\quad\quad\quad \hspace{0.65cm} \dot{u}(t)>0
\end{aligned} \right.
\\
\alpha_\mathrm{s}=\left\{ \begin{aligned}
&         6u(t)- \alpha_1 v(t-\tau), \quad  |6u(t)- \alpha_1v(t-\tau)| < 0.6
\\& 0.6, \quad \quad \quad \quad\quad \quad \hspace{0.37cm} |6u(t)- \alpha_1 v(t-\tau)| \geq 0.6
\end{aligned} \right.$

\begin{figure}[!t]
\centering
\includegraphics[width=8.4cm]{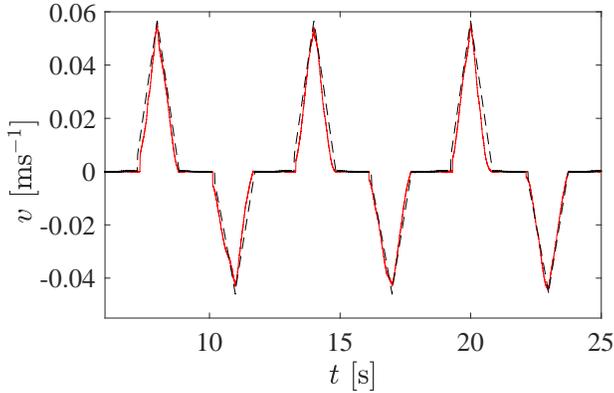}
\caption{Measured (solid) and modelled (dotted) velocity response to 6s triangular wave input with -2V, 1.5V peak to peak.}
\label{fig_sim_10}
\end{figure}
\vspace{0.25cm}
\hspace{-0.37cm}where $\tau = 3.5$ms and $v_\mathrm{cr} = 5\times 10^-6 \mathrm{ms}^{-1}$. Using this model, triangular function input with amplitude $1.5$V, $-2$V and a period of $6$s are shown in Fig. \ref{fig_sim_10}. In Fig. \ref{fig_sim_10}, the dashed line shows the results obtained by the model, and the solid line shows the results of the actual system. From the results it can be seen that the simulation results are in basic agreement with the results of the actual system, thus, it is possible to use this model to design the control system.


%

\subsection{Reference Tracking Performance}

As a comparison, PI control and the approach proposed in \cite{Xu08} are used to test the tracking performance of the piezoelectric motor to a desired reference trajectory. The parameters of the PI control are $K_\mathrm{P}=1.9\times 10^4$ and $K_\mathrm{I}=6.6\times 10^5$ which are selected based on the requirement that the minimal tracking error is obtained that does not lead to oscillatory output. The DTISMC approach proposed in \cite{Xu08} is designed based on the linearized approximation of the model (\ref{25}).
%

%

%
To proceed with the SMCPMC control law design, the sliding surface is selected as
\begin{equation}
\label{31}
s(e(t),\dot{e}(t))=e(t)+3\dot{e}(t)
\end{equation}
while the desired reference trajectory is given as $y_\mathrm{d} = 10 \left(1 + \mathrm{sin}\left(\pi t - \frac{\pi}{2}\right)\right)$ mm. According to the approach (\ref{6}), the control law is obtained as
\begin{equation}
\label{32}
u(t)=\hat{u}(t)+\frac{1}{3}\dot{e}(t)+863.1s(e(t),\dot{e}(t))+1.3\mathrm{sgn}(s(e(t),\dot{e}(t)))
\end{equation}
where the parameters $\lambda = \frac{1}{3}$, $\eta = 863.1$ and $\beta = 1.3$ are obtained based on the partial model and then tuned online. The partial model component, $\hat{u}(t)$, of the control law is given as
\vspace{0.25cm}
\begin{equation} 
\label{33}
\hat{u}(t)=\left\{ \begin{array} {lll}
\frac{\dot{v}_\mathrm{d}+\alpha_1 v_\mathrm{d}(t-\tau)+3.1023}{6},  & v_\mathrm{d}> v_\mathrm{cr} \\
\frac{\dot{v}_\mathrm{d}+\alpha_1 v_\mathrm{d}(t-\tau)+\alpha_\mathrm{s} \mathrm{sgn}(v_\mathrm{d})}{6},  & |v_\mathrm{d}|\leq v_\mathrm{cr} \\
\frac{\dot{v}_\mathrm{d}+\alpha_1 v_\mathrm{d}(t-\tau)-\alpha_\mathrm{2n}}{6}, &  v_\mathrm{d}< -v_\mathrm{cr}
\end{array} \right.
\end{equation}
\vspace{0.25cm}
\hspace{-0.12cm}and the coefficients $\alpha_1$, $\alpha_\mathrm{2n}$ and $\alpha_\mathrm{s}$ are given as
$\\  
\alpha_1=\left\{ \begin{aligned}
&         104.0154, \quad  v_\mathrm{d}(t-\tau)> 0
\\& 117.1441, \quad   v_\mathrm{d}(t-\tau)<0
\end{aligned} \right.
\vspace{0.25cm}
\\
\alpha_\mathrm{2n}=\left\{ \begin{aligned}
&         5.8216 + \left(1-e^{-30 v_\mathrm{d}(t)}\right), \quad  \dot{u}(t)\leq 0
\\& 6.8216, \quad \quad\quad\quad\quad\quad \hspace{0.57cm} \dot{u}(t)>0
\end{aligned} \right.
\vspace{0.25cm}
\\
\alpha_\mathrm{s}=\left\{ \begin{aligned}
&         6u(t)- \alpha_1 v_\mathrm{d}(t-\tau), \quad |6u(t) - \alpha_1 v_\mathrm{d}(t-\tau)| < 0.6
\\& 0.6, \quad \quad \quad \quad \quad\quad \hspace{0.5cm} |6u(t)- \alpha_1 v_\mathrm{d}(t-\tau)| \geq 0.6
\end{aligned} \right.$

\vspace{0.25cm}
\hspace{-0.37cm}Note that, due to the characteristics of the output signal, a first-order low-pass filter is used to filter the signal $\dot{e}(t)$ with a time constant of 0.1.

\begin{figure}[!b]
\centering
\includegraphics[width=8.4cm]{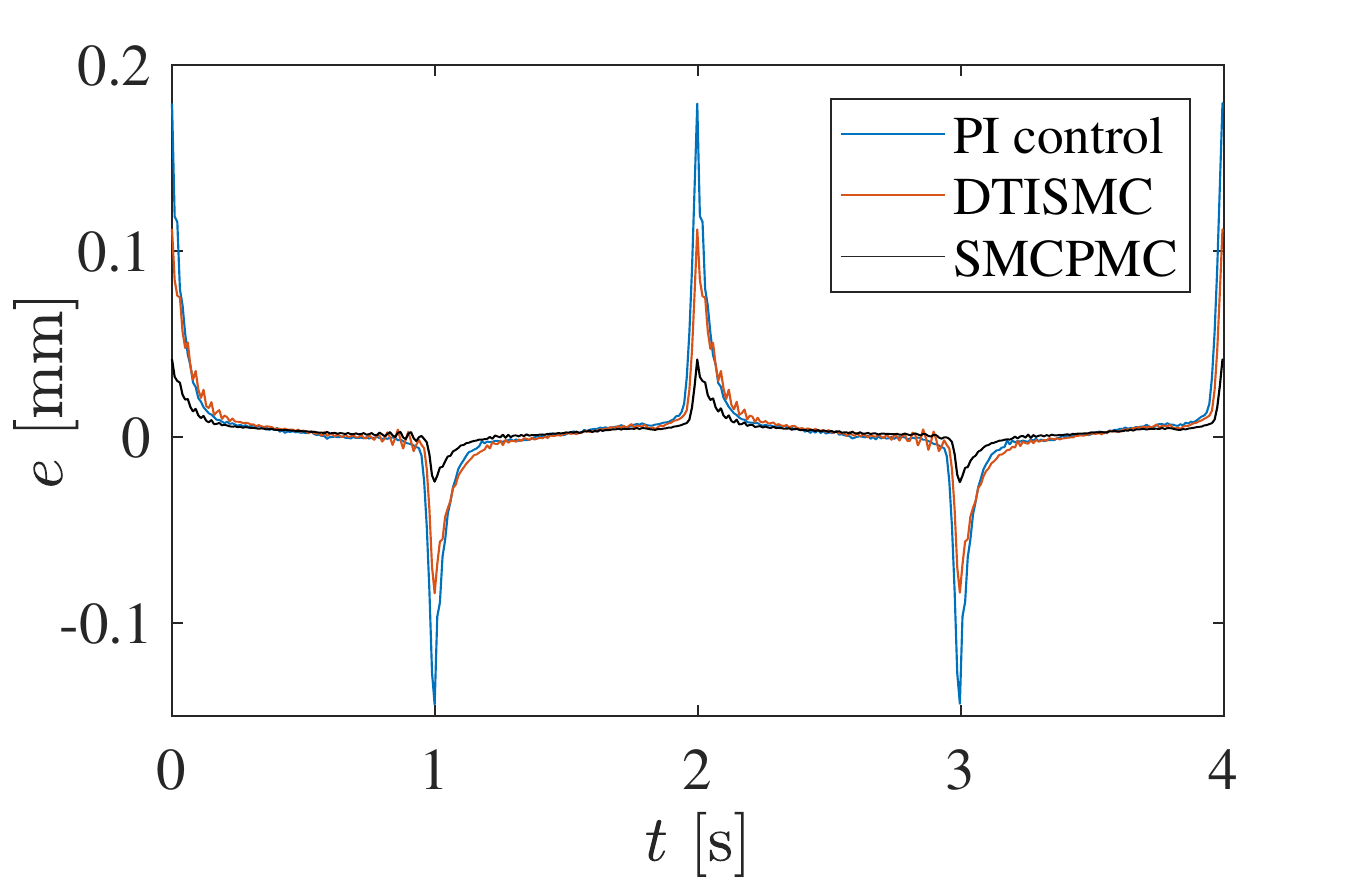}
\caption{Tracking error of PI controller, DTISMC and SMCPMC.}
\label{fig_sim_12}
\end{figure}

\begin{figure}[!b]
\centering
\includegraphics[width=8.4cm]{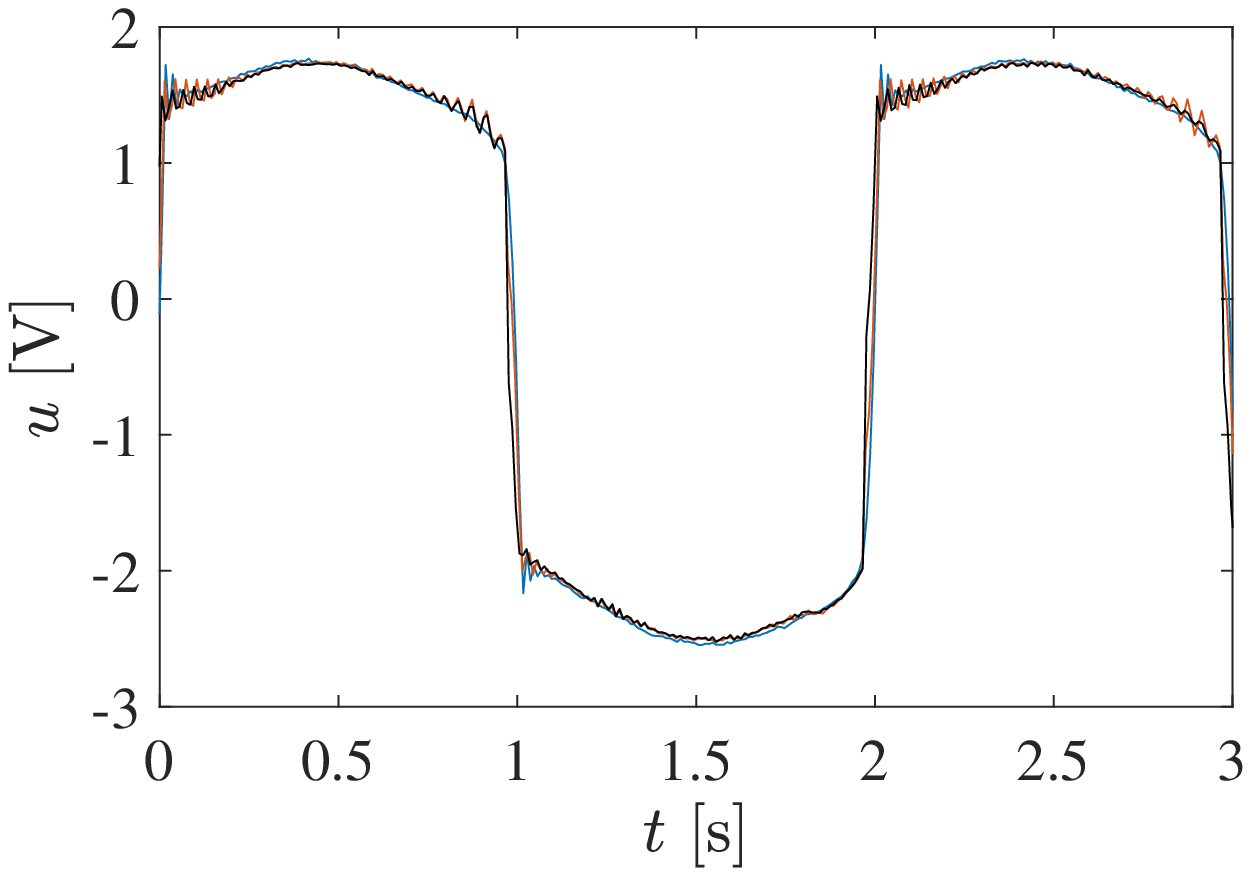}
\caption{Comparison of the control signals of PI controller (blue), DTISMC (red) and SMCPMC (black).}
\label{fig_sim_13}
\end{figure}

\begin{figure}[!t]
\centering
\includegraphics[width=8.4cm]{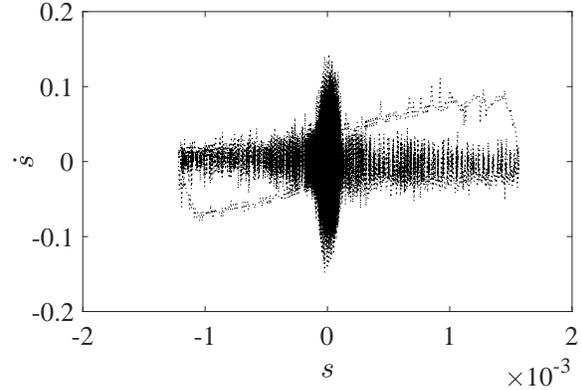}
\caption{$s(e(t),\dot{e}(t))$ versus $\dot{s}(e(t),\dot{e}(t))$ in the phase plane.}
\label{fig_sim_14}
\end{figure}

\begin{figure}[!t]
\centering
\includegraphics[width=8.4cm]{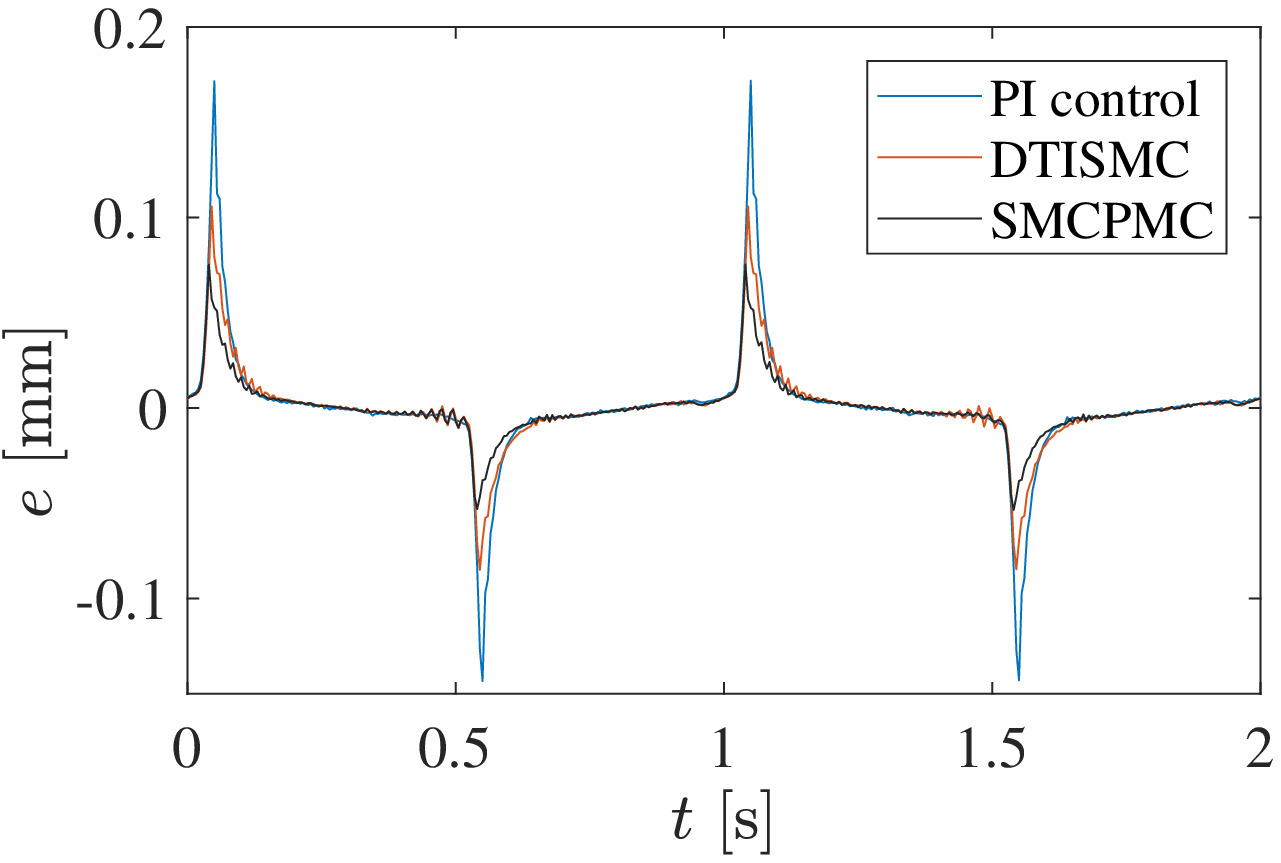}
\caption{Tracking error of PI controller, DTISMC and SMCPMC.}
\label{fig_sim_12_2}
\end{figure}

\begin{figure}[!t]
\centering
\includegraphics[width=8.4cm]{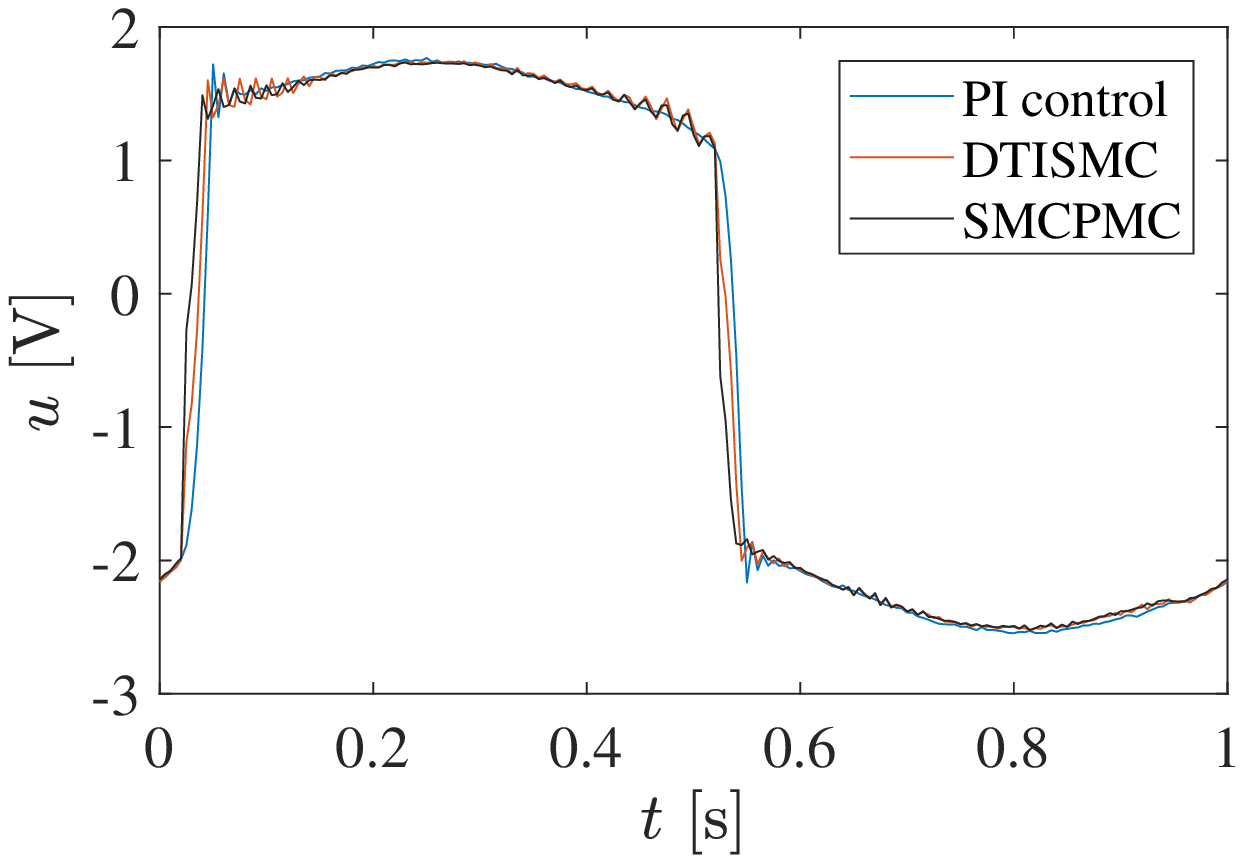}
\caption{Comparison of the control signals of PI controller (blue), DTISMC (red) and SMCPMC (black).}
\label{fig_sim_13_2}
\end{figure}

The tracking error results of all the three approaches are shown in Fig. \ref{fig_sim_12}. It can be seen that the tracking error of the SMCPMC is smaller in magnitude than that of the PI control. Even though the DTISMC approach performs better than PI control, it is unable to outperform the SMCPMC approach. This is due to the fact that the disturbance observer used in the DTISMC approach is incapable of compensating for the static friction force, however, since the partial model used for the design of the SMCPMC includes the static friction it is better able to compensate for it. The control signals of the all the three approaches are shown in Fig. \ref{fig_sim_13}. The sliding surface function $s(e(t),\dot{e}(t))$ and its derivative $\dot{s}(e(t),\dot{e}(t))$ is shown in Fig.\ref{fig_sim_14}. It can be seen that $s(e(t),\dot{e}(t))$ and $\dot{s}(e(t),\dot{e}(t))$ are well convergent in the phase plane. Finally, the speed of the reference trajectory is increased such that the new trajectory is $y_\mathrm{d} = 10 \left(1 + \mathrm{sin}\left(2\pi t - \frac{\pi}{2}\right)\right)$ mm. Using this new reference trajectory, the experiments are repeated and the results are shown in Fig. \ref{fig_sim_12_2} and Fig. \ref{fig_sim_13_2}. From the results it can be seen that the performance slightly degrades, however, the SMCPMC still outperforms the other two approaches.
\begin{remark}
Note that the control law parameters are obtained using the experimentally obtained partial model, however, those parameters need to be tuned online to improve the performance and this can be a limitation depending on the application.
\end{remark}

%

\section{Conclusion}
This work presents a SMCPMC controller based on partial modeling compensation. This approach is characterized by the full use of modeling information that is based on a model that does not need to be very accurate. Another advantage of the approach is the simplistic design and implementation process. Rigorous convergence analysis of this approach are presented while the experimental comparison with well know approaches show that the performance of SMC based approaches can be greatly improved by the addition of partial modeling compensation rather than relying on disturbance observers.

\end{document}